\documentclass{article}

\usepackage{PRIMEarxiv}

\usepackage[utf8]{inputenc} 
\usepackage[T1]{fontenc}    
\usepackage{hyperref}       
\usepackage{url}            
\usepackage{booktabs}       
\usepackage{amsfonts}       
\usepackage{nicefrac}       
\usepackage{microtype}      
\usepackage{lipsum}
\usepackage{fancyhdr}       
\usepackage{graphicx}       
\graphicspath{{media/}}     

\usepackage[english]{babel}

\usepackage{amsmath,amssymb,mathrsfs,}
\usepackage[round]{natbib}
\usepackage{braket}
\usepackage{graphicx}

\title{A crack in the track of the Hubble constant
\thanks{\textit{\underline{Citation}}: 
\textbf{Forthcoming in \textit{ Philosophy of Astrophysics},  Springer, Synthese Library, eds. N.Boyd, S. De Baerdemaeker, V. Matarese and K. Heng.}} 
}

\author{
  Marie Gueguen\\
  Institut de Physique de Rennes 1\\
  Université de Rennes 1\\
  Rennes\\
  mgueguen@uwo.ca \\
}

\begin{document}
\maketitle

\begin{abstract}
Measuring the rate at which the universe expands at a given time--the `Hubble constant'-- has been a topic of controversy since the first measure of its expansion by Edwin Hubble in the 1920's. As early as the 1970’s, Sandage et de Vaucouleurs have been arguing about the adequate methodology for such a measurement. Should astronomers focus only on their best indicators, e.g., the Cepheids, and improve the precision of this measurement based on a unique object to the best possible? Or should they “spread the risks”, i.e., multiply the indicators and methodologies before averaging over their results? Is a robust agreement across several uncertain measures, as is currently argued to defend the existence of a `Hubble crisis' more telling than a single 1 $\%$ precision measurement? This controversy, I argue, stems from a misconception of what managing the uncertainties associated with such experimental measurements require. Astrophysical measurements, such as the measure of the Hubble constant, require a methodology that permits both to reduce the known uncertainties and to track the unknown unknowns. Based on the lessons drawn from the so-called Hubble crisis, I sketch a methodological guide for identifying, quantifying and reducing uncertainties in astrophysical measurements, hoping that such a guide can not only help to re-frame the current Hubble tension, but serve as a starting point for future fruitful discussions between astrophysicists, astronomers and philosophers.
\end{abstract}

\section{Introduction}

 From the realization in the end of the 1920's by Edwin Hubble that a relation of proportionality exists between the recessional velocities of galaxies and their distance; to the crisis around the Hubble constant that currently undermines the standard model of cosmology, the history of this constant has been that of the chase of a fleeing number that kept escaping the scientists' net. 
Among the most remarkable episodes of this track: the \textit{Hubble war} in the 1970's, opposing Sandage and Vaucouleurs, arguing both about the correct methodology to adopt for measuring the Hubble constant and about its actual value\footnote{See \cite{guralp2020calibrating}.}; the dispute between Sandage and his colleague Wendy Freedman at the Carnegie Observatories of Pasadena in the 1980's, the latter defending a much higher value than the former, probably lying in the middle of the range spanned by Sandage and Vaucouleurs; the disagreement since 2014 between two opponents that nobody had seen coming: the distant and the local universe, the former with a Hubble value of $67.4 km\ s^{-1} \ Mpc^{-1}$, the latter one approaching 75, and finally the so-called and on-going Hubble `crisis' that this persisting disagreement and its apparent confirmation by the publication of new local measures in 2019 have seeded. 

The agitated history of the Hubble constant mirrors how fundamental this parameter has been for the development of our modern, precision cosmology. As one may remember from the famous words of A. Sandage, modern cosmology can be considered as the ``search of two numbers'': the values of the Hubble constant and of the $q_0$ parameter, which characterizes the deceleration in the expansion \citep{sandage1970cosmology}. But this troubled history certainly also reflects how tedious and delicate the task of measuring the Hubble constant is, and, as a result, how difficult it has been to assess the accuracy of its past measures. The determination of the Hubble constant requires one to find stellar objects a) whose luminosity is known on theoretical grounds, b) sufficiently far away from us to be freely moving (i.e., located in the so-called `Hubble flow'), and c) bright enough to be detected even that far away. But no single technique allows to  measure distance of objects satisfying all these properties. Hence, doing so necessitates to deploy a `cosmic ladder' from the nearby universe to the Hubble flow, where each of the rungs leads to a proliferation of systematic errors that must be constantly tracked and eliminated.

Yet, recent developments in astrophysics have led many scientists to consider that we know enough to consider the Hubble tension as a Hubble crisis. Such takes are grounded in the idea that 1) measurements of the Hubble constant have reached a sufficient precision for the discrepancy between early and late universe measures to meaningful, and that 2) the robustness of the high value inferred from independent late universe techniques guarantees that no systematic errors will explain away this discrepancy. Here, I argue that in the context of highly uncertain measurements, methodologies that favor tracking the unknown unknowns always have epistemic priority over robustness arguments in the sense that they constrain the appropriate domain and timing for applying such arguments.  On this basis, I contend that the current Hubble constant crisis is yet another avatar of a methodological confusion between the possible roles that different kinds of replication can play: the form of replication that robustness constitute cannot be considered as evidence of a crisis when the necessary condition of systematic replication, which promotes the track for unknown unknowns, is not successful.

Section 2 introduces the reader to the different ways of measuring the Hubble constant. In section 3, I reconstruct the reasons that have been provided to justify the idea of a Hubble crisis and how they relate to the notion of robustness. Finally, section 4 clarifies the roles that robustness and replication can play and contend that the use of robustness to establish a conclusion as dramatic as a Hubble crisis at this stage of the investigation is misguided: a day may come when a Hubble constant crisis arise and our cosmological model crashes down, but it is not this day. 

\section{How to track the Hubble constant}

Different methods have been developed since Edwin Hubble's first attempt at measuring the Hubble constant \textit{via} the cosmic ladder. One consists in inferring the Hubble constant from the early universe, for instance from the cosmic microwave background. The Santa Barbara conference of 2019 has seen new techniques based on the local universe blossom and reach a precision that makes their comparison with the most mature techniques genuinely informative. In this section, I briefly introduce some background for each of these techniques, such as to facilitate the philosophical interpretation of their concordance and of its signification for a cosmological crisis. 
\subsection{Jack and the Magic Bean: building a cosmic distance ladder in the local universe}

At first sight, measuring the Hubble constant seems quite straightforward.The expansion law that must be solved in order to measure its value takes the following form: 
\begin{equation}\label{expansion law}
c \frac{\delta\lambda}{\lambda_0}=H_0D_0
\end{equation}
where $c$ is the speed of light, $\frac{\delta\lambda}{\lambda_0}$ the redshift of the observed spectral lines of the galaxies compared to what would be expected only taking into account their distance, and $D_0$ their present distance. In other words, $H_0$ is known when the distance of an object and its redshift are known. Determining the redshift of a given stellar object is done by comparing the observed spectral lines of galaxies to the `laboratory' ones. The distance, on the other hand, is determined on the basis of two pillars: the choice of a standard candle on one hand, its apparent magnitude and its absolute magnitude on the other hand. A standard candle is an object that has a known intrinsic luminosity, referred to as its `absolute magnitude' $M$--the brightness we would measure if we were standing 10 parsec away from it\footnote{Astronomers use the notion of ``magnitude" to measure the brightness of an object. Magnitude is defined on a logarithmic scale, and the brighter an object is, the lower its magnitude. For instance, the absolute magnitude of the Sun is 4.8, but the faintest objects visible by the Hubble telescope have an apparent magnitude of 30.}. The apparent magnitude $m$ of an object corresponds to its brightness as it appears to us, taking into account its distance and the effects that interstellar dust or bright stars nearby could have on it. The distance modulus $\mu$ is equal to $m-M$ and is related to the distance $d$ in parsecs as follows: 
\begin{equation}
    \mu = 5 log_{10} d-5
\end{equation} 
The problem, as we mentioned above, is that the relevant objects for measuring the Hubble constant must be located in the `Hubble flow', and that determining the apparent magnitude of an object so far away requires a bundle of different methods that each comes with its own difficulties. There are, indeed, many phenomena  that can alter the apparent magnitude of an object along our line of sight beyond its distance. It may, for instance, appear much fainter than it should, due to the absorption of part of its spectrum by the dust surrounding it --a phenomenon referred to as `extinction'; or brighter than predicted, due to crowding effect by nearby stars. Such phenomena, among many others, must be accounted for and the apparent magnitude calibrated on this basis. Thus, objects in the Hubble flow cannot be \textit{directly} probed. They require to develop a ladder that will allow for the calibration of distances and magnitudes one rung at a time. Each rung is built on a different object, on a different technique, and on the information provided by the former rungs. Needless to say that in such a case, any error done in the first steps of the process has important repercussions on the final value found for the Hubble constant. 

 Cosmic distance ladders built to measure the Hubble constant are usually based on three rungs. The first rung `anchors' the ladder in the sense that it serves as a zero-point calibration for extinction and crowding effects. Anchors must be sufficiently close to measure their distances with geometric methods, either through trigonometric parallaxes or Detached Binary Eclipses (DEBs). Anchors usually include the Large Magellanic Cloud, the Milky Way and NGC 4258. The second rung consists of determining the distances of galaxies known as `calibrators', hosting both the selected standard candle and Type I Supernovae.  Based on the calibration done in the first rung, the difference between apparent and absolute magnitudes allows to determine the distance of galaxies hosting both our standard candle and Type I Supernovae. Ideal standard candles consist of objects whose luminosity does not depend on their mass or composition. They usually fall into one of these two categories: either stars whose luminosity varies according to a known period-luminosity law, or extremely luminous objects whose brightness is due to a well-known and well-described phenomena. They typically include, among many other examples, Cepheid stars, whose average intrinsic luminosity varies depending on the period at which they pulse, a relationship well-documented and empirically verified by Henrietta Levitt\footnote{\cite{leavitt19081777}.}. Another favored one are Type Ia Supernovae, which correspond to a rare but extremely bright explosion --around five million times the brightness of the Sun! --, that of a dying white dwarf star exceeding its critical mass. Their brightness is perfectly suited for exploring the Hubble flow. The combination of the two provide both the anchors needed for building the first rungs of the cosmic ladder, and an access to regions where galaxies are freely moving. One starts by calibrating distances to nearby Cepheids in the LMC, Milky Way or NGC 4258, before gauging distances to much farther away Cepheids. Finally, the distance of Type I Supernovae in the Hubble flow is determined, on the basis of the second rung calibration.

This picture of a ``Jack and the Magic bean" astronomer climbing the cosmic ladder to catch supernovae, as beautiful as it is, is however anything but simple. As we mentioned above, each rung comes with many traps and errors propagates from one rung to the others. Maybe surprisingly, the zero-point calibration of the ladder is one of the trickiest part of the process and the largest source of systematic uncertainties. Uncertainties associated with the anchors carry over as systematic errors and have a huge possible impact on the determination of $H_0$. Yet, these sources of uncertainties are not only important, but impossible to reduce otherwise than by improving the accuracy of observational tools. Among them, one can include the distance to these anchors, extinction, but also difficulties related to converting the I-band photometric system of space telescopes to ground-based telescopes\footnote{See \cite{freedman2019carnegie}, p.11.}, both needed. The second major source of uncertainties comes from the fact that it is usually difficult to find a statistically significant sample of galaxies that host both the relevant standard candles and Type I Supernovae. In the case of the Cepheids, four decades or research have allowed to build a sample of only 19 host galaxies\footnote{\cite{riess2022comprehensive} have succeeded in more than doubling this sample in 2022, with a Hubble value now at $72.53 \pm 0.99 km\ s^{-1}\ Mpc^{-1}$.}. But more generally speaking, standard candles are rarely really 'standard', as their luminosity may actually depend on their age or metallicity\footnote{In astronomy, the metallicity of a star corresponds to the heavy elements it contains, a `heavy' element being any element other than hydrogen or helium.}. Random velocities of specific galaxies can be perturbed by local gravitational perturbations, thereby complicating the task of determining their redshift if the statistical sample of standard candles is not big enough to average away these perturbations. In sum, each rung comes with its ensemble of systematic and statistical proliferating errors, each of which could significantly distort the final value inferred for $H_0$. Significant progresses have yet allowed to improve the precision of the measure of the Hubble constant based on Cepheids up to 1\%\footnote{\cite{riess2019large}.}, after many years of rigorous investigation to explore and reduce the systematics associated with this technique.

\subsection{Hubble constant in the early universe}

What characterize measures of the Hubble constant based on the primordial universe is their model-dependence. One cannot infer the value of the Hubble constant from the early universe without already assuming a cosmological model. This feature holds both for measures inferred from the cosmic microwave background (henceforth CMB) or from the Baryon Acoustic Oscillation (BAO). For space reasons, I chose to limit the introduction to the early universe measures of $H_0$ to the CMB measure, but a detailed and accessible introduction to the BAO determination of $H_0$ can be found for instance in \cite{Fong2011MeasuringTH}. 

A couple words on the CMB first. When the primordial universe got cold enough for the first neutral hydrogen atoms to form --the epoch of 'recombination', photons decoupled from matter and started to free-stream across the universe. These photons have been propagating ever since, and the relic of this radiation is what is referred to as the `cosmic microwave background' (henceforth CMB). This fossil electromagnetic radiation offers an extraordinary window into the early universe, as it provides a map of how matter was distributed across the universe at the time of decoupling and thus informs us about fundamental parameters of the $\lambda$CDM model, included its matter density $\Omega_{m}$. Assuming the standard cosmological model, and the dark energy density $\Omega_{\lambda}$ and spatial curvature $k$ that characterise this model, secondary parameters such as the Hubble constant can be derived through the following equation:
\begin{equation}
H_z= H_0\sqrt{\Omega_m(1+z)^3}+\Omega_{\lambda}+\Omega_k(1+z)^2
\end{equation}

A nearly exact geometrical degeneracy exists however, both for $\Omega_{\lambda}$ and $k$, that make different cosmological models based on different $\Omega_{\lambda}$ and spatial curvature $k$ compatible with the anisotropies mapped by the CMB\footnote{See e.g. \cite{Fong2011MeasuringTH} or \cite{efstathiou2020lockdown}. The Planck 2018 results released in \cite{collaboration2020planck} seem to possibly break this degeneracy however.}. Hence the need and importance of a model-independent measure of $H_0$ that could waive this degeneracy, such as measures in the local universe.

\section{A tale of two values: the Hubble crisis}
 Now, here lies the problem: the results delivered by the two techniques, the one based on the early universe and the Cepheids-based one, do not agree. In other words, the expansion of the universe as measured from our local universe is much faster than that predicted on the basis of the CMB, by almost 8\%.  The difference between the two is significant, close to $5\sigma$: the value announced by the $SH_0ES$ team in 2019, led by A. Reiss and working with Cepheids, was $74.03\pm1.42km\ s^{-1}\ Mpc^{-1}$\footnote{See \cite{riess2019large}.}, when the value obtained from the CMB after the last release of Planck results\footnote{See \cite{collaboration2020planck}.} is $67.4\pm0.5km\ s^{-1}\ Mpc^{-1}$. Until now, this difference was not considered as too alarming. As we saw, the measure of the Hubble constant using the cosmic ladder with Cepheids as standard candles is a delicate task. Cepheids are young stars, thereby living in the dusty and crowded center of galaxies--an environment that maximizes extinction and crowding effects, whose period-luminosity depends on their age and metallicity, and whose nature itself is a problem, inasmuch as variable stars necessitate many exposures during several observational campaigns, adding new sources of systematic errors to account for. No wonder then that the first measure of $H_0$, based on Cepheids, was off by an order of magnitude: Edwin Hubble estimated the constant value around $500km\ s^{-1}\ Mpc^{-1}$.  The complexity of the first technique, the multiple systematic and statistical errors that could affect each rung of the ladder, and the degeneracy associated with $H_0$ in the early universe context could have legitimately let us think that, as the accuracy of the measures improve in the future, the results would have converged, especially as, as figure 1 shows, these values did progress a lot over the last two decades, and globally toward a possible convergence. 

\begin{figure}
\includegraphics[scale=0.7]{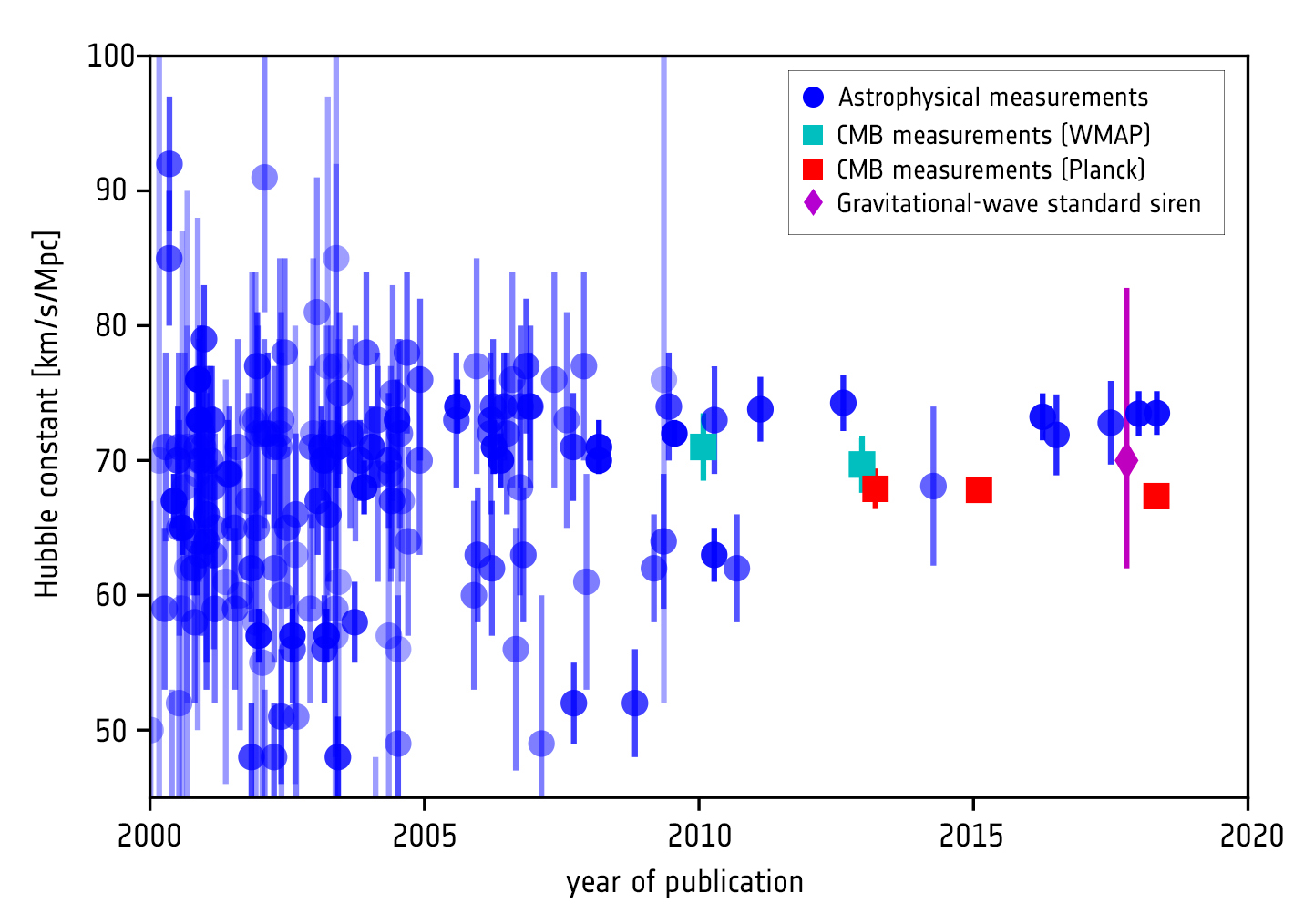}
\caption{Evolution of measurements of the rate of the Universe’s expansion over the past two decades. `Astrophysical measurements' refer to the cosmic ladder based on Cepheids. Credit: @ESA}
\end{figure}

The conference that was held in Santa Barbara in 2019 has however shaken this confidence and consolidated the gap that one was hoping to bridge. Over the last decade, many new techniques have indeed been developed to measure the Hubble constant independently of the Cepheids and the CMB, in order to break the tension between the two--especially as it becomes less and less clear how the precision of these measurements could further be improved. Four new measures were released during the conference that did not resolve the tension, but on the contrary corroborated the high value found by the $SH_0ES$ team, turning an apparent disagreement into a genuine problem, and generating a strong feeling of crisis among cosmologists\footnote{A nice overview of the different reactions heard during this conference can be found in the paper ``Cosmologists debate how fast the universe is expanding"( https://www.quantamagazine.org/print), written for \textit{Quantamagazine} by N. Wolchover.}. Part of the goal of this paper is to elucidate the reasons that underlies such a feeling and whether these reasons are justified. 
\subsection{The blossoming of new measurement techniques}
The main reason driving this sense of crisis is the fact that these new techniques are considered as independent measurements, that is, measurements that differ enough from the Cepheids measure to exclude possible common sources of systematic errors. Let us briefly review three of these techniques\footnote{I leave aside the result obtained by the Surface Brightness Fluctuations method, whose error bar is so large that its agreement/disagreement with other results is not meaningful. See more on this in \cite{potter2018calibrating}. } to assess the extent to which this claim is justified. 
\begin{itemize}
    \item $H_0liCOW $:  $H_0=73.3\pm1.7 \pm1.8km\ s^{-1}\ Mpc^{-1}$\\
The $H_0licow$\footnote{See \cite{wong2020h0licow}.} project uses strong gravitational lensing, i.e., the distortion of spacetime produced by supermassive objects, to measure the Hubble constant. This lensing phenomena permits to obtain multiple images of a same object, based on the different paths that the electromagnetic radiation follows given the curvature of spacetime. The idea of $H_0licow$ is to study the light emitted by 5 quasars, i.e, extremely luminous active galactic nuclei whose magnitude varies. Massive galaxies between us and these objects act as magnifying and distorting lens that multiple the images of the lensed target. Since light takes a different path for each of the images, the oscillation of the luminosity is delayed for each of these pictures. Thus, given that the distance travelled by light depends on the expansion of the universe, the time delay between each image allows to calculate the Hubble constant.
     \item	Mira variables\footnote{See \cite{huang2020hubble}.} : $H_0=73.3\pm3.9 km\ s^{-1}\ Mpc^{-1}$\\
This method is a variant of the cosmic distance ladder, but using red pulsating stars called MIRAs as standard candles. The main sequence of the life of a star consists of converting the hydrogen contained in its heart in helium through nuclear fusion. At some point of the life of the star, after hydrogen and helium in the core are fully exhausted, the fusion starts in the outer shell which will expand, sometimes up to 1 AU; then cool down and shrink, before expanding again. This cycle is what makes MIRAs-type of stars good standard candle, because this period-luminosity relationship is fully determined by the mass and radius of the star. Note that this cycle, that lasts at minimum 100 days, is even longer than the Cepheids',  whose fluctuations usually range from 1 to around 50 days.
\item 	The Megamasers Cosmology Project\footnote{See \cite{pesce2020geometric}.}: $H_0=73.9\pm3.0 km\ s^{-1}\ Mpc^{-1}$\\
Megamasers and gravitational lensing are especially interesting techniques, as both offer the possibility of a direct measure of the Hubble constant, skipping the rungs of the ladder altogether. The characteristic of interest of masers is the equivalent of the laser effect in the microwave domain. The rough idea is the following: the spontaneous emission of a photon generated by an atom's transition to its fundamental state triggers a cascade of similar emission. The incident photon provokes the deexcitation of another atom; and thus the emission of another photon. Hence, the initial photon is so to speak photocopied up to a very powerful electromagnetic beam. Megamasers are typically located around 1pc of the center of a galaxy, close to the active galactic nuclei that can stimulate the surrounding gas clumps or water clumps (in the cases of water masers). Hydrogen and oxygen atoms composing the water maser absorb the galactic nuclei energy and radiate it in the form of a microwave 22 Hz beam that can be detected by Very Long Baseline Interferometry (VLBI). From this stable radiation can be inferred the velocities of gas clumps and water clouds orbiting the nuclei, their radius from the nuclei and distance to the galaxy, and their host galaxy's redshift.
\end{itemize}

These new techniques are all very promising, but also very recent. Beyond the many known sources of errors and unknown unknowns to uncover associated with them, their youth comes with a high price, that of relying on limited (and so possibly biased) statistical sample. Take for instance the $H_0licow$ or the Megamaser Cosmology Project results: the former was based on only 7 lensed quasars in their 2019 paper, and the latter had only 4 megamasers-- including NGC4258 which is used as an anchor for cosmic ladder techniques. One has to grant however that they are based on totally different physical assumptions, a fact that renders very unlikely the possibility of unveiling a common plausible source of errors explaining their convergence toward a high value. 

This makes the fourth result published during the Santa Barbara conference even more disturbing.  Indeed, the Carnegie-Chicago-Hubble program, led by W. Freedman, presented their Hubble value during this exact same conference, based on a new version of the cosmic ladder using the Tip of the Red Giant Branch stars (TRGBs) as a standard candle. Their announcement, far from solving the issue, created a new puzzle, since their value lies right in the middle between the low value based on the CMB and the high value of the local universe methods, at $H_0=69.8\pm 0.8km\  s^{1}\ Mpc^{-1}$.
\subsection{Houston, we have a rogue measure}
TRGBs offers an excellent standard candle to calculate the distance of supernovae. This phase of the evolution of red giants corresponds to the moment when a star of around 1 to $2M_{\odot}$ has exhausted the hydrogen of its core and started the fusion in the outer shell. Unlike what happens for stars of higher mass though, the core does not contract but is entirely sustained by the electron degeneracy pressure. As a result, the temperature increases in the core as the helium piles up there, without any corresponding dilatation of the star, until the temperature reaches 100000K. At this point a triple $\alpha$ reaction is triggered. Under the combined effects of the temperature and the pressure exerted in the core, the fusion of helium into beryllium and into carbon turns into a runaway reaction and creates an extremely violent flash of helium, with a release of energy superior to the entire output of a whole galaxy. The infrared luminosity of stars going through an helium flash is independent from their mass and composition, which makes them eligible to the title of standard candles. 

Probably one of the most interesting features of this technique is to constitute the perfect counterpart to the Cepheids technique. This means that, even though we currently have no way to decide whether one of these results, that obtained based on the Cepheids or that based on TRGBS, is  correct, a comparison between them is both very telling and very informative, as each one fills the lacuna of the other. Whereas Cepheids are short lived-stars that live in dusty environments, TRGBs are old stars living in isolation, in the outskirts of galaxies. As such, they are not as exposed to extinction and crowding effects as Cepheids are. Likewise, while the period-luminosity relationship of the latter depend on their metallicity, that of TRGBs can be accurately accounted for in two different ways: first, the infrared I-passband is almost not affected by their metallicity. Second, this metallicity manifests itself in the color of the star, by a widening of the RGB color that has been well-studied and calibrated empirically. Finally, TRGBs are not variable stars and do not require multiple exposures.  As we will see in the next section, these complementary features consists an ideal investigation path and offers a fecund scenario not only to test the robustness of the Hubble value, but also to discover new sources of uncertainties not necessarily accounted for in the report of the accuracy of these measurements. The question is thus the following: how can we explain the fact that, TRGBs excluded, the different methods based on the local universe agree on a high value and the methods based on the early universe on a low value, whereas at the same time the two methods that are the closest to each other, the more complementary and the more likely to agree fail to do so? How can we account for this success on one hand and this failure on the other?

\section{Should we call it a crisis?}
 How should we thus react to these diverging measures? For many cosmologists, as reflected in the number of papers attempting to resolve the Hubble problem published since 2019, the convergence of the $SH_0ES$, the $H_0licow$, the Miras and the Megamaser techniques towards a high value of $H_0$ is taken to indicate that the standard model of cosmology is undergoing a crisis. Although it is difficult to reconstruct the exact argument supporting a crisis, the papers that endorse the idea of a `Hubble tension' or a `Hubble crisis' tend to agree on the following statements: first, the discrepancy is much higher than the error estimates associated with each of these results. Put differently, the error bar is sufficiently small for a meaningful assessment of the discrepancy. Second, the independence of the techniques converging toward a high value for $H_0$ excludes that one single, shared, source of systematic errors could waive the tension. Hence, the latter will likely resist an improvement of the accuracy of these techniques and a subsequent decrease in the size of their error bar \footnote{See notably \cite{verde2019tensions}, \cite{efstathiou2020lockdown}, \cite{riess2020expansion}.}. The quote below summarizes this view:

\begin{quote}
Given the size of the discrepancy and the independence of routes seeing it, a single systematic error cannot be the explanation. [...] Moreover, a suite of low redshift, different, truly independent measurements, affected by completely different possible systematics, agree with each other; it seems improbable that completely independent systematic errors affect all these measurements by shifting them all by about the same amount and in the same direction" (\cite{verde2019tensions}, p.7).\footnote{See also \cite{riess2020expansion}, p.2).}    
\end{quote}

\subsection{From robustness to reliability}
 Although the word is never explicitly stated in the discussion, this line of argument captures what constitutes the core of robustness analysis as theorized within the tradition starting with \cite{levins1966strategy}, \cite{levins1993response} and \cite{wimsatt2012robustness}\footnote{See also \cite{weisberg2006robustness} and \cite{weisberg2012simulation}for a more recent take on this version of robustness analysis. Of course, many schools of thoughts have arised since Levins' and Wimsatt's accounts of robustness analysis. The goal here, however, is not to address whether robustness is a sound tool, but whether the kind of robustness allegedly displayed in the Hubble context supports the existence of a crisis. Therefore, the paper focuses on their version of robustness, which seems to capture the line of argument defended by cosmologists.For a more up-to-date account of how robustness is used in the actual practice of scientists, see among many others \cite{soler2012characterizing}.}. Robustness analysis has been famously suggested by Levins as a way to assess the trustworthiness of models in the absence of a background theory providing analytically soluble equations. Since models have to be simplified to get predictions susceptible to be measured against nature, a method must be developed in order to evaluate the impact of these simplifications on the predictions of the model and to determine ``whether a a result depends on the essentials of the model or on the details of the simplifying assumptions'' (\citealt{levins1966strategy}, 423). One way to do so, to Levins' eyes, is to compare different models $M_1$, $M_2$, ..., $M_N$ of the same target system, where each model is conceived as the intersection of a common, plausible core C and of an unshared, variable part $V_1$, $V_2$, ..., $V_N$, and to look for a connection between C and a predicted property R (\citep{levins1993response}. The plausible core includes the biological or physical assumptions that are undergoing the test, while the unshared part corresponds to different idealizations or simplifications used to make the problem tractable. If one can show that the intersection of C with the union of the $V_i$ implies R, then one can establish under certain conditions\footnote{For instance, the condition that the set of $V_i$ exhausts the space of admissible possibilities.} that C alone implies R--put differently, that R does not depend on the $V_i$, but one the common core of all models whose adequacy is under test:
\begin{quote}
(...)Thus the search for robustness as understood here is a valid strategy for separating conclusions that depend on the common [...] core of a model from the simplifications, distortions and omissions introduced to facilitate the analysis, and for arriving at the implications of partial truths.”
 (\cite{levins1993response}, 554).
\end{quote}
Levins, however, remained rather careful about what can be learned through robustness; at least in his 1993 piece: 
\begin{quote}
    (...) the more inclusive the set of $V_i$'s, the more we can have confidence that C implies R. If we feel that the set of $V_i$'s spans a wide enough range of possibilities, then we may generalize to claim that C usually implies R, a result that is not very exciting as a mathematical theorem but may be good biology(\cite{levins1993response}, 554).
\end{quote}
It was \citeauthor{wimsatt2012robustness} in 1981 who tightened the bond between robustness and reliability. According to him, robustness analysis is defined through the following three principles: first, it is a procedure aiming at distinguishing the ``reliable from the unreliable", second, it requires one to show the invariance of that which reliability is scrutinized over independent\footnote{For space reasons, I will not dwell on how to characterize what `independence' means here, given that models have to be of the same target and thus presumably share some core assumptions to be even comparable. This term is present in both Wimsatt's and Levin's work, but never fully elucidated and is subject to controversy. See for instance \cite{schupbach2015robustness}.} processes or models, in order to build confidence in their independence from these; and finally to determine the scope of this invariance. Hence, within this framework, establishing reliability is no longer a mere possible goal of robustness, but one of its core and definitional tenets--robustness and reliability go hand by hand, and where one is to be found, the other is expected: 
\begin{quote}
[A]ll the variants and uses of robustness have a common theme in the distinguishing of the real from the illusory; the reliable from the unreliable; the objective from the subjective; the object of focus from artifacts of perspective; and, in general, that which is regarded as ontologically and epistemologically trustworthy and valuable from that which is unreliable, ungeneralizable, worthless, and fleeting. (p. 46) 
\end{quote}

Since Wimsatt, robustness has been generally accepted as an indicator of reliability, that is, as evidence that a prediction is not an artifact of specific modelling assumptions\footnote{See for instance \cite{weisberg2006robustness}, \cite{weisberg2008robust} or \cite{soler2012characterizing}, but also \cite{parker2011climate} for a criticism.}.\citeauthor{orzack1993critical} have however convincingly undermined this claim, in a beautiful paper that seeded many of the questions about robustness that have been debated ever since. As they see it, there are three possible scenarios resulting from a Levins-like robustness reasoning: 
\begin{itemize}
    \item Scenario 1: We already know that one of the M's among $M_1$, $M_2$, ..., $M_N$ is true\footnote{Although we do not endorse the terminology in terms of 'truth' that Orzack and Sober use, we will keep it here in order to remain faithful to the authors.}. In that case, if for all $i$, $M_i$ implies R, then R must be true. As emphasized in \cite{justus2012elusive} (p.797), this inference is unproblematic but also relatively uninteresting, given that robustness is precisely needed in those cases where no observations or no analytic solutions is available that could establish the truth of one of the $M_i$.
    \item Scenario 2: We known that all the models are false. In this case, we have no reasons to believe that the fact that each $M_i$ implies R is evidence that R is true. Their simple example illustrates this point beautifully: if all models we compare in population biology admits natural selection as the only force acting on the size of the population, then all models will predict population with infinite size. That they agree does not say anything about the truth of the prediction, but only about the convenience of the assumption (\cite{orzack1993critical}, p.538).
    \item Scenario 3: We do not know whether one of the models is true. This is the most common situation in astrophysics and cosmology, and robustness is precisely used in these contexts to help to establish the reliability of predictions converging across different models. According to Orzack and Sober, we have no more reasons yet in this situation to infer that R is true that we have in the second scenario: "if we do not known whether one of the models is true, then it is again unclear why a joint prediction should be regarded as true (ibid, p.538-539).
\end{itemize}

From a purely inferential point of view, I think that Orzack and Sober make a valid logical point here in emphasizing that we have no reason to consider R as true in any of the last two scenarios. Nonetheless, I am still willing to grant that the last scenario can correspond to very different epistemic situations, and that for each of them the degree of confidence possibly supported by the robustness of R or its value as a heuristic guide could vary a lot. A comparison where recently developed techniques, with many shared assumptions, and from which little is understood, are compared, does not support a high degree of confidence in R. But it seems reasonable to say that a much higher degree of confidence in R would be justified--in the words of Levins, would be ``good astrophysics"-- if the comparison is performed across mature and independent models, that have been rigorously examined and studied such that systematics and statistical errors have been identified and reduced to the best of our knowledge. The question is thus the following: which one of this situation corresponds to the Hubble constant crisis's scenario? Are we in the epistemic position to apply robustness and infer conclusions with a high degree of confidence on this basis? Or are we putting the cart before the horse?

\subsection{Temporary discrepancy vs. residual discrepancy }
When are we justified in thinking that a discrepancy is symptomatic of a crisis? My --presumably uncontroversial-- answer to this question would be: when the discrepancy is not a temporary one, but a \textit{residual}\footnote{I am borrowing this term from a talk delivered by Jim Weatherall on June, $30^{th}$ 2022 at the History, Philosophy and Sociology of Cosmology and Astroparticle Physics organized in Bonn, entitled: ``Two tooth fairies and a dentist: Closing the Loop on $\Lambda CDM$".}
discrepancy. That is, when we found ourselves in the case where known sources of systematic and statistical errors have been quantified and sufficiently reduced for a comparison between the two values thus obtained and their error bars to be significant, and when enough efforts have been done to chase unknown unknowns, i.e., new sources of yet undiscovered errors.  A \textit{residual} discrepancy, in other words, is the discrepancy that remains after adequate efforts have been deployed to identify, quantify and reduce all possible sources of uncertainties that could explain away the disagreement. 

Of course, this does not mean that we need to be absolutely certain that all sources of errors have been excluded. It does mean, however, that the significance of the tension is directly related to the assumption that all sources of uncertainty have been identified and accounted for. If we have strong evidence that there are still errors, which are not accounted for but which could resolve the tension, the robustness of the tension does not have sufficient epistemic strength to justify a call for a crisis. 

Up to the Santa Barbara conference, one can see that the discrepancy between the early and the late universe measures was interpreted as a temporary one, thus merely seen as a tension rather than as a crisis. What changed after the Santa Barbara conference is that the robustness of the values resulting from the local universe measurements\footnote{This robustness holds only at the price of excluding the TRGBs' result of course. The $SH_0ES$ team \citep{yuan2019consistent}  has justified such an exclusion on the basis of, as they argued, a calibration error on the TRGBs side. This claim has now been debunked several times (\cite{freedman2020calibration}, \cite{freedman2021measurements}, \cite{mortsell2021hubble}.} was interpreted as an indication that the discrepancy had gone from temporary to residual. Now, do we have good reasons on the basis of the new results to think that the epistemic situation switched from temporary to residual? And is robustness the appropriate tool to decide whether this is the case, that is, if systematic and statistical errors have been sufficiently purged from our measurements? 

In the remainder of the paper, I content that we have clear evidence that not only the discrepancy is not residual, but that robustness has so far been unsuccessful in detecting unknown systematic errors in our case study, notably because of the emphasis on comparing measurements \textit{as independent as possible}.  I illustrate the latter point by showing that the robustness of the high late-universe value is blind to the systematics since acknowledged (notably) in time-delay cosmography. Next, I show that the tool that could diagnose these errors is actually in competition with robustness and thus often neglected despite its epistemic priority.

\subsubsection{The example of time-delay cosmography }
Announced in 2019, the $H_0liCOW$ result was considered as one of the most important evidence of a Hubble crisis. As mentioned above, a measure based on gravitational lensing is a \textit{direct} measure of the Hubble constant, in that it does not require to appeal to the ladder technique. Furthermore, the $H_0liCOW$ and the Cepheids' measurements are as independent as two measures of the Hubble constant can be. They rely on completely different objects and different physics, whereas the Mira project is a version of the cosmic ladder which might suffer from the same issues as the Cepheids or the TRGB technique (due to common anchors for instance) and the Megamaser Project involves the maser located in NGC4258, which is a common zero-point for the Cepheids and the TRGB measures. Thus, an agreement between the two independent measures released by $H_0liCOW$ and by $SH_0ES$ was considered as particularly exciting and telling, and a major reason to interpret the  the Hubble tension as a Hubble crisis. 

But this technique is a really young one, and a lot of work still needs to be done to understand how the different assumptions that enters this measurement might distort the result. Note that this is not pure speculation about possible future developments for time-delay measurements of gravitational lensing. The effects of relaxing assumptions about the mass density profile of the deflector have already been carefully studied, with surprising results. Indeed, one of the most important sources of uncertainties in time-delay lensing is the mass profile of the deflector. If no assumptions is made about it, the precision of the measurement, based on the 7 $H_0liCOW$ lenses, drop from 2\% precision to 8\%. Such an error budget is far too important to resolve the Hubble tension. So where do the mass assumptions used in this context come from?

A lens model should ideally be able to reproduced the observables associated with the lens with as few unconstrained parameters as possible. With respect to quasars astrometry however, there are too few observational constraints to reach this standard, which means that different models can reproduce the same set of observations but give different $H_0\delta t$ product and thus different values for $H_0$. This degeneracy can be broken  by relying on stellar kinematics--from which the above 8\% uncertainties estimate is obtained-- or, for an improved precision, by further constraining the mass distribution of the lensing system. Traditionally, the two main solutions adopted  are that of a power-law, or of a constant mass-to-light ratio plus the so-called Navarro-Frenk-White dark matter halo density profile\footnote{Note that the NFW profile is challenged, as it fails to reproduce the observations especially for low-surface-brigtness galaxies. See for instance \cite{bullock2017small}.} inferred from simulations (\cite{navarro1997universal}). These mass assumptions are not however chosen on theoretical grounds or observational constraints. To be sure, surveys did show that the mean slope of the density profile of lenses is nearly isothermal. But this slope is an average, and thus need not be adequately described by a power law (\cite{schneider2013mass}. On the contrary, there are good reasons to think that the central regions of the lens would significantly depart from a perfect power law. Yet, the studies published by \cite{birrer2020tdcosmo} and \cite{birrer2021tdcosmo} have shown that, with a sample of lenses  increased from 7 to the 33 lenses from the SLAC- TDCOSMO collaboration, if the mass modelling assumptions are relaxed to be maximally degenerated, then the value obtained for the Hubble constant is no longer of $H_0=73.3\pm1.7 \pm1.8 km\ s^{-1}\ Mpc^{-1}$, but of $H_0=67.4\pm 4.1 \pm 3.2 km\ s^{-1}\ Mpc^{-1}$, no longer in significant tension with the CMB measure. Such a result demonstrates how much we need to further improve our understanding of the systematics for time-delay cosmography before we can claim a 2\% precision measurement that does not involve unjustified assumptions when it comes to gravitational lensing. Clearly, the high Hubble value with a small error bar highly depends on an assumption that has no strong physical justification.  Acknowledging what we still do not know, while not taking any bets, leads to a low value with a much larger error bar. But more importantly, it shows that the robustness of the high Hubble value is no guarantee that this high value is not an artefact from systematic errors. Had the robustness argument for a crisis been taken at face value, the track for these unknown facts would have stopped and the importance of the mass distribution assumption not properly understood in improving the precision of the measurement.

\subsubsection{Systematic replication and unknown unknowns}
Now, what about the Cepheids's claim of 1\% precision measurement, grounded in more than four decades of refining the ladder technique? Can we legitimately believe that new systematics could remove the discrepancy at this stage of the scientific investigation? 

Before addressing this question, a short detour through another toolbox, that of replication, is necessary. The Replication Crisis\footnote{An introduction to the replication crisis and to its importance can be found in \cite{romero2019philosophy}.}, according to which many findings in social, behavioral and biomedical sciences have failed to replicate at alarming rates, has led to many interesting developments when it comes to understand what is a replication and what purpose it can serve. I will briefly present a typology of replication adapted from \cite{schmidt2016crisis}, \cite{schmidt2016shall}, \cite{zwaan2018making} and \cite{fletcher2021role} before going back to our current issue. The typology I suggest decline replication along four categories. It is important to note that these categories are better conceived as covering a spectrum and revealing different aspects of replication than as clean-cut separations between different types of replication\footnote{One way to think about this spectrum is that the typology aims to capture situations going from: nothing is changed in the replication, only one thing is changed, to several if not all are changed. How you quantify the number of variables that vary also depends on how fine-grained your perspective is.}: 
\begin{itemize}
    \item \textit{Direct replication}: direct replication is an attempt to reproduce exactly the original study, on a different statistical set. It is especially useful to excludes errors related to the statistical sample or to contextual factors. In the case of the Cepheids' technique for instance, a successful direct replication would consist of obtaining the same Hubble constant value based on different Cepheid calibrators. 
    \item \textit{Methodological replication}: this kind of replication is a simple re-analysis of an experiment, ideally by another team. As \citeauthor{fletcher2021role} puts it, methodological replication "ensures that the results of a scientific study are not due to data-entry, programming or other suchlike technical errors", or in general to human errors. The re-analysis of the $SH_OES$ data found in \cite{javanmardi2021inspecting} is such an example, among many ones, of such methodological replication. 
    \item \textit{Systematic replication}: it consists in systematically varying one of the variables of the experiment or measure while maintaining the others fixed.  The goal of systematic replication is to help to identify which variables causally contribute to the final outcome, but more importantly to better understand and circumscribe the causal contribution of a given variable to the result. Contrary to the other three, systematic replication is more about understanding a protocol, measurement, experiment or model than about assessing the reliability of its prediction. 
    \item \textit{Conceptual replication}: here, the goal is to measure the same phenomenon or test the same hypothesis as in the initial study, but by using different methods, techniques or models. Conceptual replication subsumes robustness analysis: an agreement between different models or different types of measurements amount to a successful replication. But given that the more independent the measurements under comparison are, the stronger the link between robustness and reliability, robustness is on the far-end of the spectrum--it is a form of conceptual replication that insists on the fact that the original and the replicated study ideally would have nothing in common but the targeted value. In other words, the most extreme form of conceptual replication is needed to warrant strong robustness-based conclusions. The alleged agreement between the four local measures  detailed above is supposedly of such nature: it is a successful conceptual replication, allegedly between fully independent measures, inasmuch as the TRGB result is excluded. 
    
\end{itemize}

The comparison between the TRGB and the Cepheids results exemplifies how the replication spectrum goes from systematic replication to conceptual replication. As the reader may remember, TRGB and Cepheids have complementary weaknesses and strengths, as well as common  and independent zero-point anchors: on the Cepheids side, the zero-point calibration is based on Milky Way paralloax, on the LMC or on NGC4258, and the sample of Type I SNe used is the Supercal sample. On the TRGB side, the zero-point calibration has been done on the basis of the distance modulus to the LMC based on DEBs + Hubble Space Telescope parallax calibration in \citeyear{freedman2019carnegie}, and on the basis of the LMC, NGC 4258 and the Milky Way  globular clusters in \citeyear{freedman2020calibration}. The sample of supernovaes can be that of the Carnegie-Chicago Program or the SuperCal SNIa sample used by the $SH_0ES$ team. Galaxies where the calibrators (either TRBG or Cepheids) can be found  as well as SNe include 18 host galaxies on the TRGB side, 19 on the Cepheids's side, 11 of which are common to the two groups. In other words, the comparison between the two results can be constructed such as to maximally overlap and leave only the choice of standard candle as the variable explored--which amounts to the perfect picture of a systematic replication, or to be fully independent, which would allow in the case of an agreement to a perfect conceptual replication, inasmuch as the standard candle used is no longer considered a mere variable but a method. As it turns out, neither the systematic replication nor the conceptual replication are successful replications in this context: recently updated TRGB and Cepheids measurements result in differing values of $H_0 = 69.6\pm 1.9 km\ s^{-1}\ Mpc^{-1}$ \citep{freedman2019carnegie} for the TRGB and $73.04 \pm 1.4$ \citep{riess2020expansion} for the Cepheids. But the conceptual replication does not teach us anything about how to locate the problem, as differences in several variables does not allow to pin down the most probable culprit. Differences in one variable only, as it the case with systematic replication, does not fall prey of this problem. If the two measurements only differ in the sample of supernovae chosen, then the calibration or a possible bias in the sample of Supernovae is most likely responsible for the disagreement. It is only because the Carnegie Chicago Hubble Program led by W.Freedman proceeded to a detailed systematic comparison between TRGB stars and Cepheids that we now have a better idea about where to look for possible unidentified unknown unknowns. While the two methods show excellent agreement on the distance modulus to 28 galaxies for instance, the study shows that this agreement no longer holds when comparing the distance to the 10 SNIa host galaxies that the two have in common. Future observational campaigns with much higher resolution, notably thanks to the JWST telescope, might be in a position to elucidate this disagreement.

The failure of systematic replication not only indicates with no possible doubts that new systematics are yet to be discovered, but informs us about where to look for them: if one wants to test an hypothesis about a possible source of systematics (e.g., the distance modulus to the LMC), the overlap between these two techniques easily allows to design a crucial test that permits to verify such an hypothesis-- for instance by comparing the TRGB result to the value inferred from Cepheids only on the basis of the Milky Way and of  NGC 4258. Likewise, the fact that the metallicity can be constrained for TRGB stars allows to decouple the problem of metallicity and of extinction, given that the TRGB I-band is not affected by metallicity effects. Metallicity and extinction can thus be individually solved,  and the measure of the extinction obtained from TRGB stars can inform the calibration of Cepheids for common objects\footnote{For a detailed comparison between TRGB stars and Cepheids for the different replications performed, see section 4 of \cite{freedman2019carnegie}.}. Hence the epistemic superiority of systematic replication in context of highly uncertain measurements, and the need to wait for successful systematic replication before applying robustness arguments: 
\begin{itemize}
    \item Given that the focus of robustness is on comparing models or measurements that are as independent as possible, arguments drawn on its basis remains mute and offer no explanations and no guide to locate the problem when the robustness analysis fails. A good example of this is a comparison between the success of the robustness strategy when excluding the TRGB result and its failure when including it. Once the claim of poor calibration on the TRGB side is excluded, how to account for this failure? How can scientists decide where to start to explain it?  Systematic replication, on the other hand, is maximally informative and is in position to identify possible sources of failure. Conceptual replication does not have in principle to be mute about such possibles sources of errors, but the part of the conceptual replication spectrum that corresponds to robustness does, inasmuch as it necessitates to maximize the independence of different measurements. 
    \item As we have seen above, a successful robustness analysis or conceptual replication does not logically establish the reliability of a given prediction. But the failure of systematic replication demonstrates unequivocally that new systematics have yet to be identified. Hence, if we grant the claim made in 4.1 that a high degree of confidence is better supported when the comparison holds between mature and independent models, free as much as possible of unknown systematic and statistical errors,  then systematic replication does not only have the epistemic priority to assess a discrepancy, but also the chronological priority. Indeed, it is the role of systematic replication to diagnose whether such unknown unknowns have still to be identified. It is only when successful systematic replications are performed that one can be confident that the most important sources of systematic and statistical errors have been accounted for. In other words, if systematic replication cannot be performed successfully, it demonstrates that the conditions for applying conceptual replication understood as robustness are not yet met, at least in the conditions that would allow a high degree of confidence in conclusions drawn from it\footnote{Although space reasons prevent me to expand on this point, it would be interesting to analyze the different scenarios that can arise: failure of CR vs success of SR, failure of both, and so on and so forth. Here we only address the claim that a specific kind of conceptual replication, that usually referred to as robustness analysis, is sufficient to warrant the reliability of the late-universe measurements and thus the problematic nature of the discrepancy.}. In the case of the Hubble constant, the failure of the systematic replication performed on the Cepheids and TRGB results show that the precision of these measurements, though by far the most mature techniques for determining the value of the Hubble constant, has not reached a sufficient level for robustness arguments to be telling and/or trusted. 
\end{itemize}

 Two remarks are needed to qualify the claim made here. First, one does not have to reject robustness analysis altogether on these grounds. Robustness analysis and systematic replication are complimentary tools and can very well work together to address different problems. They cannot however be deployed at the same stage of the inquiry, as the latter indicates when the conditions for justified inferences based on robustness are met. The appeal to robustness has no epistemic grounds if systematic replication is not successfully achieved. If systematic replication fails, no meaningful robustness-based conclusions about the reliability of the measure can be drawn. Second, when used too early, robustness analysis is actually an obstacle to its companion and leads to neglect the chase for unknown unknowns. This happens because robustness, as an extreme form of conceptual replication, focuses on developing independent techniques that have, ideally, absolutely nothing to do with each other--e.g., time-delay cosmography and Cepheids-based distance ladder. On the other hand, systematic replication requires measurements techniques sufficiently close to each other to be mutually informative, as Cepheids and TRGB can be. Robustness deployed too early leads to developing techniques that are too independent from each other to offer the grounds needed for systematic replication. Hence the need to understand their roles and places in the scientific investigation, so as to not let robustness becomes the crack in the track of the Hubble constant. The failure of systematic replication offered by the TRGB and the Cepheids' measurements tells us that the discrepancy between early universe and local universe measurements is not a residual discrepancy, contrary to what the defenders of a crisis would like to see on the basis of their robustness strategy. Moreover, it informs us about how measurements can be further improved and where to start for doing so. Robustness cannot be used to justify such as crisis if it is not established that the track for unknown unknowns has not gone far enough, and that we are indeed comparing mature and well-understood techniques. There might be a cosmological crisis to come, but such a crisis is not justified by the current epistemic situation, and certainly not on the basis of robustness arguments made at this stage of the scientific investigation.

\section{Conclusion}
Recent developments in astrophysics have seen the community working on the Hubble constant shaken by the robustness of the high value found by local universe measurements, and subsequently by the significant discrepancy between this value and the one obtained from early universe measurements. Some have gone as far as claiming that these developments proves that the standard model of cosmology is undergoing a crisis.  The robustness of the high-values, they contend, shows that the discrepancy between the early and late universe will hold, and thus is the long-wished for evidence that new physics is needed to amend the standard model. I hope to have shown here that we do not have good reasons to interpret the discrepancy as residual on the basis of robustness arguments, and to think that we are currently facing a crisis. The Hubble debate does not offer the conditions that would warrant a strong degree of confidence in robustness-based inferences. The fact that we seem to have reached 1\% precision measurements is not yet a sign that systematic errors have been almost eliminated, but that the unknown systematics are getting harder and harder to track. The blossoming of new techniques is an opportunity not to establish the robustness of the discrepancy, but to use their overlap to deploy systematic replication and refine our understanding of where the skeletons could still be hiding.

\bibliographystyle{chicagoa}
\bibliography{hubble.bib}

\section*{Acknowledgments}
The author is very grateful to Chris Smeenk, Barry Madore and Wendy Freedman for their helpful feedback. They would also like to thank the Bonn History and Philosophy of Physics Seminar, the APC colloquium (Paris 7), and the audience of the "What can Astrophysics teach us about Reblicability?" workshop organized by the CSH center (Bern) for insightful discussions. Finally, the author would like to thank the two anonymous referees for their helpful feedback. This paper is based on work done while funded  under the John Templeton Foundation grant: “New Directions in Philosophy of Cosmology” (grant number 61048).

\end{document}